\documentclass[11pt,twoside]{article}

\usepackage{amsmath,amsfonts,amssymb,amsthm}

\newcommand{\field}[1]{\mathbb{#1}}

\newcommand{\R}{\field{R}}
\newcommand{\C}{\field{C}}

\newcommand{\F}{\mathcal{F}}
\newcommand{\Fin}{\mathcal{F}_{\rm fin}}
\renewcommand{\H}{\mathcal{H}}
\newcommand{\h}{\mathfrak{h}}
\renewcommand{\L}{\mathbf{B}}
\newcommand{\D}{\mathcal{D}}

\newcommand{\dGamma}{\mathrm{d}\Gamma}

\newcommand{\eps}{\varepsilon}
\newcommand{\ph}{\varphi}

\newcommand{\curl}{\mathrm{curl}}

\newcommand{\restricted}{|\grave{}\,}

\newcommand{\expect}[1]{\mbox{$\langle #1 \rangle $}}         
\newcommand{\sprod}[2]{\left\langle #1,#2 \right\rangle}       

\newcommand{\supp}{\operatorname{supp}}

\newcommand{\Rea}{\operatorname{Re}}

\newtheorem{theorem}{Theorem}
\newtheorem{lemma}[theorem]{Lemma}

\begin{document}
\title{\bf Exponential Decay and Ionization Thresholds in Non-Relativistic
Quantum Electrodynamics}

\author{ {\vspace{5pt} M. Griesemer\footnote{Work partially supported by U.S. National
Science Foundation grant DMS 01-00160.} } \\
\vspace{-4pt}\small{Department of Mathematics, University of
Alabama at Birmingham,} \\
\small{Birmingham, AL 35294, USA}\\ }

\date{}

\maketitle

\begin{abstract}
Spatial localization of the electrons of an atom or molecule is
studied in models of non-relativistic matter coupled to quantized
radiation. We give two definitions of the ionization threshold.
One in terms of spectral data of cluster Hamiltonians, and one in
terms of minimal energies of non-localized states. We show that
these two definitions agree, and that the electrons described by a
state with energy below the ionization threshold are localized in
a small neighborhood of the nuclei with a probability that
approaches 1 exponentially fast with increasing radius of the
neighborhood. The latter result is derived from a new, general
result on exponential decay tailored to fit our problem, but
applicable to many non-relativistic quantum systems outside
quantum electrodynamics as well.
\end{abstract}

\section{Introduction}

If an atom or molecule is in a state with total energy below the
ionization threshold, then all electrons are well localized near
the nuclei. In non-relativistic quantum mechanics this finds its
mathematical expression in the discreteness of the energy spectrum
below the ionization threshold and in the exponential decay of the
corresponding eigenfunctions. When the electrons are coupled to
the quantized radiation field, then there is no discrete spectrum
anymore and the ground state is the only stationary state
\cite{BFS1, BFSS}. Nevertheless, all states in the spectral
subspace of energies below the ionization threshold are
exponentially well localized as functions of the electron
coordinates. To prove this is the main purpose of this paper.
Localization of the electrons below the ionization threshold is
necessary to justify the dipole approximation \cite{BFS1}, and it
plays an important role in proving existence of a ground state
\cite{BFS1,BFS3,GLL} and for Rayleigh scattering \cite{FGS2}.

The ionization threshold is the least energy that an atom or
molecule can achieve in a state where one or more electrons have
been moved ``infinitely far away"  from the nuclei. To give a more
precise definition we need a mathematical model for atoms and
molecules. A (pure) state of $N$ electrons and an arbitrary number
of transversal photons shall be described by a vector in the
Hilbert space \(\H_N=\H_{\mathrm el}\otimes\F\), where
$\H_{\mathrm el}$ is the antisymmetric tensor product of $N$
copies of $L^2(\R^3;\C^2)$, appropriate for $N$ spin-$1/2$
fermions, and $\F$ is the bosonic Fock space over
$L^2(\R^3,\C^2;dk)$. The nuclei are static, point-like particles
without spin. Let $H_N$ denote the Hamilton operator  generating
the time evolution in $\H_N$, and let $H_N^0$ be the same
Hamiltonian without external potentials (nuclei). We assume that
the dynamics of the electrons is non-relativistic and that the
forces between material particles (electrons and nuclei) drop off
to zero with increasing distance. In view of the latter assumption
a natural definition for the ionization threshold $\tau(H_N)$ is
\begin{equation}\label{int:tau}
  \tau(H_N) := \min_{N'\geq 1} \{E_{N-N'}+ E_{N'}^0\},
\end{equation}
where \(E_{N-N'}=\inf\sigma(H_{N-N'})\),
\(E_{N'}^0=\inf\sigma(H_{N'}^{0})\), and $E_{N=0}=0$. Let $m$ be the mass of the electron and
let \(|x|=(\sum_{j=1}^n x_j^2)^{1/2}\) for \(x\in\R^n\).
We prove that, for all real numbers $\lambda$ and $\beta$ with
\(\lambda+\beta^2/(2m)<\tau(H_N)\),
\begin{equation}\label{int:result}
  \Big\|e^{\beta|x|}E_{\lambda}(H_N)\Big\| < \infty,
\end{equation}
and that in states with energy above $\tau(H_N)$ the electrons
will not be localized in general. Thus $\tau(H_N)$ is in fact a
threshold energy separating localized from non-localized states.
The question of whether the binding energy \(\tau(H_N)- E_N\)
is positive or not, is not addressed in this paper, see
however \cite{GLL}.

Our proof of \eqref{int:result} consists of two independent parts.
First we give an alternative definition of the ionization
threshold which better captures the idea of a localization
threshold, and we prove exponential decay below it. Then we show
that the two definitions agree.

The alternative definition is as follows. Let \(D_R=\{\ph\in
D(H)|\,\ph(x)=0,\ \text{if}\ |x|<R\}\), and define a threshold
energy $\Sigma(H_N)$ by
\begin{equation}\label{int:Sigma}
  \Sigma(H_N) = \lim_{R\to\infty}\left(\inf_{\ph\in D_R,\,
  \|\ph\|=1}\sprod{\ph}{H_N\ph}\right).
\end{equation}
Delocalization above $\Sigma(H_N)$ is obvious, and localization
below $\Sigma(H_N)$ will be derived from the only assumptions
that $H_N$ is self-adjoint, bounded from below, and that
\begin{equation}\label{int:com}
  [[H_N,f],f] = -2|\nabla f|^2
\end{equation}
for all bounded smooth functions $f(x)$ with bounded first
derivatives. The latter assumption is satisfied for the positive
Laplacian $(-\Delta)$, and hence for all operators $-\Delta + I$
with $[[I,f],f]=0$. Examples include the commonly traded models of
non-relativistic atoms coupled to quantized radiation, as well as
many Schr\"odinger operators outside quantum electrodynamics.

The second part of the proof, that \(\tau(H_N)=\Sigma(H_N)\), is
the hard part. The inequality \(\tau(H_N)\leq\Sigma(H_N)\)
requires localizing both the electrons and the photons, and in
particular their field energy. This was done in \cite{GLL}. To
show that \(\tau(H_N)\geq \Sigma(H_N)\) we construct suitable
(compactly supported) minimizers $\ph_0$ and $\ph_{\infty}^R$ of
$H_{N-N'}$ and $H_{N'}^0$, respectively, where $\ph_{\infty}^R$ is
localized at a distance $R$ from the origin. We than merge these
states into a single state $\psi_R\in \H_N$. The problem is to do
this in such a way that
\(\sprod{\psi_R}{H_N\psi_R}=\sprod{\ph_0}{H_{N-N'}\ph_0}+
\sprod{\ph_{\infty}^R}{H_{N'}^0\ph_{\infty}^R}+o(1)\) as $R\to
\infty$.

In the context of QED the first result of the form
\eqref{int:result} is due to Bach, Fr\"ohlich and Sigal
\cite{BFS1}, who proved exponential binding for small coupling and
away from the ionization threshold of $H_N$ with \emph{zero}
coupling. The threshold energy $\tau(H_N)$ was introduced in
\cite{GLL} where it was shown that $E_N$ is an eigenvalue of $H_N$
if $\tau(H_N)> E_N$. The paper \cite{GLL} also contains an easy
argument showing that \emph{eigenvectors} of $H_N$ with
eigenvalues below $\tau(H_N)$ exhibit the exponential decay
implied by \eqref{int:result}. For $N$-particle Schr\"odinger
operators the ionization threshold defined by the analog of
\eqref{int:tau} is the least point of the essential spectrum. This
is known as the HVZ-Theorem \cite{HunSig}. That the analog of
\eqref{int:Sigma} also characterizes the beginning of the
essential spectrum is a result due to Arne Persson \cite{Persson}.
Exponential decay for $N$-body eigenfunctions with discrete energy
was first proved by O'Conner \cite{Connor}. See Agmon's book
\cite{Agmon} for more results on the exponential decay of
solutions of second order elliptic equations.

Section~\ref{sec:abs} contains the general theorem on exponential
decay in an abstract Hilbert space setting. In
Section~\ref{sec:qed} this result is applied to quantum
electrodynamics and the main result on equality of the thresholds
is formulated. Its proof is given in Section~\ref{sec:proofs}. The
Appendix collects technical results and notations used in the
proofs.

\section{The Abstract Argument}\label{sec:abs}

In this section $q: D\times D \to \C$ denotes a densely defined,
closable, quadratic form that is bounded from below and defined on
a domain $D\subset\H$ in a Hilbert space $\H$. We assume that $\H$
is a closed subspace of a Hilbert space \(L^2(\R^n)\otimes \F\)
and that $\H$ is invariant with respect to multiplication with
bounded (measurable) functions that depend on $|x|,\ x\in\R^n$,
only. Here $\F$ is an arbitrary, additional Hilbert space. In our
applications $\F$ will be the tensor product of spin and Fock
space and $\H$ the subspace with the symmetry required by the
nature of the particles.

On the quadratic form $q$ we make the further assumption, that for
each $f\in C^{\infty}(\R^n;\R)$ with \(f, \nabla f \in
L^{\infty}(\R^n)\) and with $f(x)=f(|x|)$, there exist constants
$a$ and $b$ such that
\begin{eqnarray*}
(i) && fD\subset D\\
(ii) && |q(f\ph,f\ph)|\leq a q(\ph,\ph)+ b
\sprod{\ph}{\ph}\\
(iii) && q(f^2\ph,\ph) + q(\ph,f^2\ph) -2 q(f\ph,f\ph) =
-2\sprod{\ph}{|\nabla f|^2\ph}
\end{eqnarray*}
for all \(\ph \in D\). Requirements (i) and (ii) are mild
technical assumptions which ensure that property (iii) extends to
all $\ph$ in the domain of the closure of $q$. Equation (iii) is
the basis of the so called IMS (localization) formula for
Schr\"odinger operators. To verify it for a quadratic form $q$
that is defined by a symmetric operator $H : D\subset \H\to \H$ it
is useful to know that \(f^2 H + H f^2 -2f H f = [[H,f],f]\).
Assumption $(iii)$ then becomes
\[  [[ H,f],f] = -2|\nabla f|^2,\]
which holds for the positive Laplacian $(-\Delta)$ and hence for
all operators $-\Delta+I$ in $\H$ with $[[I,f],f]=0$. Some
examples, other than those in the next section, are
\(H=(-i\nabla+A(x))^2+V(x)\) with a classical vector potential
$A(x)$ and scalar potential $V(x)$ (choose $\F=\C$), and
Schr{\"o}dinger operators with restricted domains
$\Omega\subset\R^n$ (\(\H=L^2(\Omega)\subset L^2(\R^n)\otimes
\C\)), or with potentials that are constant away from a strip, as
in wave guides defined by potential wells.

Given $R>0$ let \(D_{R}=\{\ph\in D: \ph(x)=0\ \text{for}\ |x|<
R\}\) and define
\begin{gather}\label{sigma-bound}
\Sigma_R = \inf_{\ph\in D_{R},\,\|\ph\|=1}
q(\ph,\ph)\quad\text{and}\quad \Sigma =\lim_{R\to \infty}
\Sigma_{R}.
\end{gather}
The numbers $\Sigma_R$ are finite because $q$ is bounded from
below and because, by (i), $D_R$ is not empty. But $\Sigma$ may
take on the value $+\infty$.

\begin{theorem}[Exponential decay]\label{thm:main1}
  Suppose the quadratic form $q$ introduced above satisfies the assumptions
  (i), (ii), and (iii), and let $H$ denote the unique self-adjoint operator associated
  with the closure of the form $q$. If $\lambda$ and $\beta$ are
  real numbers with $\lambda+\beta^2<\Sigma$, then
  \[ \Big\|e^{\beta|x|} E_{\lambda}(H)\Big\|<\infty,\]
  where $E_{\lambda}(H)$ is the resolution of the identity for $H$.
\end{theorem}

\emph{Remarks.} (1) For Schr\"odinger operators \(-\Delta+V\) on open
domains $\Omega\subset\R^n$ with Dirichlet boundary conditions
and with \(V_{-}\ll -\Delta\) the above theorem implies that the
spectrum below $\Sigma$ is discrete. In fact
\((-\Delta+1)^{-1/2}e^{-\beta|x|}\) is compact and hence so is
\(E_{\lambda}(H)=
E_{\lambda}(H)(-\Delta+1)^{1/2}\,(-\Delta+1)^{-1/2} e^{-\beta|x|}\,
e^{\beta|x|}E_{\lambda}(H)\)
for \(\lambda+\beta^2<\Sigma\).

(2) Everything in this section holds equally for any norm $|x|$ on
$\R^n$ that is induced by an inner product $x\cdot y$, provided
that $\Delta$ is used to denote the Laplace-Beltrami operator with
respect to the metric $g(x,y)=x\cdot y$.
\\

The following proof is inspired by the proof of binding in Bach et
al.\cite{BFS1}.

\begin{proof}
Let $Q(H)\subset\H$ denote the form domain of $H$, i.e., the
domain of the closure of $q$. We use $q$ to denote the closure of
$q$ as well. $Q(H)$ is the closure of $D$ with respect to the
form norm $\|\cdot\|_q$ associated with $q$. By assumptions (i)
and (ii), multiplication with a bounded function $f\in C^{\infty}(\R^n)$ with
bounded derivatives is a bounded linear operator on
$(D,\|\cdot\|_q)$ and hence extends to a bounded linear operator
on $(Q(H),\|\cdot\|_q)$. In particular
\begin{equation}\label{eq:abs1}
 f Q(H) \subset Q(H)
\end{equation}
and (iii) extends from $D$ to $Q(H)$.

Let $E=\inf\sigma(H)$. We may assume $\Sigma>E$, for otherwise
there is nothing to prove. Let $\chi_{2R}$ denote the
characteristic function of the set \(\{x\in\R^n:|x|\leq 2R\}\). We
first show that
\begin{equation}\label{eq:abs2}
H_R:= H+(\Sigma_R-E) \chi_{2R} \geq \Sigma_R - \frac{C}{R^2}
\end{equation}
for all $R$ with $\Sigma_R\geq E$ and some constant $C$. Pick
\(j_1,j_2\in C^{\infty}(\R_{+})\) with \(j_1^2+j_2^2\equiv 1\),
\(\supp (j_1)\subset \{t\leq 2\}\) and \(\supp(j_2)\subset \{t\geq
1\}\). Let \(j_{i,R}(x)=j_i(|x|/R)\). Then by \eqref{eq:abs1} and
since (iii) holds on $Q(H)$,
\begin{equation}\label{eq:IMS}
\begin{split}
  H_R &= \frac{1}{2}\sum_{i=1}^2\big( j_{i,R}^2 H_R + H_R
  j_{i,R}^2  \big)\\
  &= \sum_{i=1}^2 j_{i,R} H_R j_{i,R} - \sum_{i=1}^2 |\nabla
  j_{i,R}|^2
\end{split}
\end{equation}
in the sense of forms on $Q(H)$. By definition of $\Sigma_R$ and
by the construction of $j_{i,R}$,
\begin{equation*}
\begin{aligned}
j_{1,R}\, H_R\, j_{1,R} &= j_{1,R} (H + \Sigma_R-E) j_{1,R} \geq \Sigma_R j_{1,R}^2\\
j_{2,R}\, H_R\, j_{2,R} &\geq j_{2,R} H j_{2,R} \geq \Sigma_R
j_{2,R}^2.
\end{aligned}
\end{equation*}
Hence \eqref{eq:abs2} follows from \eqref{eq:IMS} and from
$|\nabla j_{i,R}|=O(R^{-1})$.

Let \(\Delta:=[\inf\sigma(H),\lambda]\), where
$\lambda+\beta^2<\Sigma$, and pick $R\in\R$ so large that
\(\lambda+\beta^2<\Sigma_R-C/R^2\). This $R$ is kept fixed in the
following. Let \(\delta := \Sigma_R-C/R^2 - \beta^2-\lambda
> 0\), and choose a function \(g_{\Delta}\in C_0^{\infty}(\R;[0,1])\)
such that $g_{\Delta}\equiv 1$ on $\Delta$ and
\(\supp(g_{\Delta})\subset (-\infty,\lambda+\delta/2]\). Then, by
\eqref{eq:abs2}, $g_{\Delta}(H_R)=0$ and therefore
\begin{equation}\label{eq:abs3}
  g_{\Delta}(H) = g_{\Delta}(H)- g_{\Delta}(H_R).
\end{equation}
We now show that \(e^{\beta|x|}(g_{\Delta}(H)- g_{\Delta}(H_R))\)
is bounded. To this end, we define
\[  f(x) := \frac{\beta\expect{x}}{1+\eps\expect{x}},\qquad \expect{x}=(1+|x|^2)^{1/2}\]
and show that, \(e^{f}(g_{\Delta}(H)- g_{\Delta}(H_R))\) is
bounded uniformly in $\eps>0$. Note that $f\in C^{\infty}(\R^n)$,
is bounded and that $|\nabla f|\leq \beta$. Let $\tilde
g_{\Delta}$ be the almost analytic extension \(\tilde
g_{\Delta}(x+iy) = (g_{\Delta}(x)+iyg_{\Delta}'(x))\gamma(y)\)
where $\gamma\in C_0^{\infty}(\R)$ equals one in a neighborhood
of $y=0$. By the almost analytic functional calculus (see
\cite{Dav})
\[ g_{\Delta}(H) = -\frac{1}{\pi}\int
\frac{\partial \tilde g}{\partial \bar{z}} (z-H)^{-1}\, dx\, dy\]
and hence, using \eqref{eq:abs3} and a resolvent identity we can
write
\[ e^f g_{\Delta}(H) = \frac{1}{\pi}\int
\frac{\partial \tilde g}{\partial \bar{z}} e^f
(z-H_{R})^{-1}e^{-f} e^{f} (\Sigma_R-E)\chi_{2R} (z-H)^{-1}\, dx\,
dy\] whose norm we estimate from above as
\begin{eqnarray*}
  \|e^f g_{\Delta}(H)\| &\leq& \sup_{z\in \supp(\tilde g)} \|
e^f(z-H_{R})^{-1}e^{-f}\|\, \|e^f\chi_{2R}\|_{\infty}(\Sigma_R-E)\\
&& \times \frac{1}{\pi}\int \left|\frac{\partial \tilde g}{\partial
\bar{z}}\right| \|(z-H)^{-1}\| \, dx\, dy.
\end{eqnarray*}
The norm $\|e^f\chi_{2R}\|_{\infty}$ is bounded uniformly in
$\eps>0$ and the integral is finite. To estimate
$\|e^f(z-H_{R})^{-1}e^{-f}\|$ let \(H_{R,f}:= e^f H_R e^{-f}\)
with domain \(D(H_{R,f})= e^f D(H)\) and note that
\[ (z-H_{R,f})^{-1} = e^f(z-H_R)^{-1}e^{-f}\]
as can easily be seen by direct computation. In particular, the
resolvent sets $\rho(H_{R,f})$ and $\rho(H_R)$ coincide. Let
\(\ph\in D(H_{R,f})\subset Q(H)\) and $\|\ph\|=1$. Then
\begin{eqnarray*}
2\Rea\sprod{\ph}{H_{R,f}\ph} &=& \sprod{\ph}{(e^f H_R
e^{-f}+ e^{-f} H_R e^{f})\ph}\\
&=& \sprod{\ph}{e^{-f}(e^{2f} H_R + H_R e^{2f})e^{-f}\ph}\\
&=& 2\sprod{\ph}{(H_{R}-|\nabla f|^2)\ph}
\end{eqnarray*}
where $(iii)$ was used in the last equation. In conjunction with
\eqref{eq:abs2} this shows that, for $z\in \supp(\tilde g)$,
\begin{equation}
\Rea\sprod{\ph}{(H_{R,f}-z)\ph} \geq \Sigma_R-C/R^2 -\beta^2 -
\Rea(z) \geq \delta/2
\end{equation}
and hence that $\|(H_{R,f}-z)\ph\|\geq \delta/2\|\ph\|$. Since
\(\rho(H_{R,f})=\rho(H_R)\supset \supp(\tilde g)\), it follows
that
\[\|(z-H_{R,f})^{-1} \| \leq 2/\delta\] for $z\in \supp(\tilde
g)$, which completes the proof.
\end{proof}

\section{Atoms Coupled to Quantized Radiation}\label{sec:qed}

In this section we apply the abstract result of the previous
section to systems of $N$ charged, non-relativistic quantum
particles, interacting with the quantized radiation field. Since
we are mainly interested in the case of electrons in the field of
static nuclei, the bulk of the exposition deals with this case. At the
end we comment on the more general case of particles from
different species.

In the ``standard model" of non-relativistic QED the Hilbert space
of a system of $N$ electrons and an arbitrary number of
transversal photons is the tensor product
\[ \H_N = \wedge_{i=1}^N L^2(\R^3;\C^2) \otimes \F_{f}\]
of the antisymmetric product of $N$ copies of $L^2(\R^3;\C^2)$,
appropriate for $N$ spin-1/2 fermions, and the bosonic Fock space
\(\F_{f}=\oplus_{n\geq 0}\otimes_{s}^n L^2(\R^3,dk;\C^2)\), where
the factor $\C^2$ accounts for the two possible polarizations of
the transversal photons. Let $\D_N\subset\H_N$ be the subspace of
sequences $\ph=(\ph_0,\ph_1,\ldots)$ where
\[ \ph_n \in C_{0,a}^{\infty}((\R^3\times\{1,2\})^N;\C)\otimes
\otimes_s^n L^2_0(\R^3,\C^2)  \] and $\ph_n=0$ for all but
finitely many $n$. The index $a$ indicates that the functions are
antisymmetric with respect to permutations of the $N$ arguments,
and $L^2_0(\R^3;\C^2)$ is the space of compactly supported
$L^2$-functions. Clearly $\D_N$ is a dense subspace of $\H_N$.

The Hamilton operator \(\tilde H_N : \D_N\subset\H_N\to\H_N\) of
our system is given by
\begin{equation}\label{eq:ham}
\tilde H_N = \sum_{j=1}^N (p_j+\sqrt{\alpha}A(x_j))^2 +
\frac{g}{2}\sqrt{\alpha}\sigma_j\cdot B(x_j) +V + H_f,
\end{equation}
where $p_j=-i\nabla_{x_j}$, $A(x_j)$ is the quantized vector
potential in Coulomb gauge evaluated at the point $x_j$,
$B(x_j)=\curl A(x_j)$ is the magnetic field, $\sigma_j$ the
triple of Pauli matrices
\((\sigma_j^{(1)},\sigma_{j}^{(2)},\sigma_{j}^{(3)})\) acting on
the spin degrees of freedom of the $j$th particle, $V$ is a real-valued
potential, and $H_f$ is the Hamilton operator of the field
energy. The parameter $\alpha$ is the fine structure constant and
the coupling constant $g\in\R$ is arbitrary, to allow for a
simultaneous treatment of the interesting cases $g=2$ and $g=0$.

Formally $A(x)$ is given by
\begin{equation*}
  A(x) =  \sum_{\lambda=1,2}\int_{|k|\leq
\Lambda} \frac{1}{\sqrt{|k|}} \eps_{\lambda}(k) \Big[ e^{ik\cdot
x} a_{\lambda}(k) + e^{-ik\cdot x} a_{\lambda}^{*}(k) \Big]d^3k
\end{equation*}
where $\Lambda<\infty$ is an arbitrary but fixed ultraviolett
cutoff. For every $k\neq 0$ the two polarization vectors
$\eps_{\lambda}(k)\in\R^3,\ \lambda=1,2$ are normalized, orthogonal to
$k$ and to each other.

The operators $a_{\lambda}(k)$ and $a_{\lambda}^{*}(k)$ are the
usual annihilation and creation operators, satisfying the
canonical commutation relations
\begin{equation*}
  [a_{\lambda}(k),a_{\mu}^{*}(k')] =
  \delta_{\lambda\mu}\delta(k-k'),\qquad
  [a_{\lambda}^{\#}(k),a_{\mu}^{\#}(k')]=0.
\end{equation*}
In terms of $a_{\lambda}(k)$ and $a_{\lambda}^{*}(k)$ the field
Hamiltonian is given by
\begin{equation*}
  H_f = \sum_{\lambda=1,2}\int d^3k\,|k|
  a_{\lambda}^{*}(k)a_{\lambda}(k).
\end{equation*}
See Appendix \ref{sec:Fock} for mathematically more proper definitions of $A(x)$ and
$H_f$.

The potential $V$ is the sum of the external potential and the
Coulomb two-body potentials for each pair of electrons. For the
purpose of the results to be proved in this section, however, it
suffices to assume that
\begin{equation*}
  (H1) \qquad V\in L^2_{\rm loc}(\R^{3N};\R),\ \text{and}\ V_{-}
  \leq \eps (-\Delta) + C_{\eps}\ \text{for all}\
\eps>0,
\end{equation*}
and, of course, that $V$ is symmetric with respect to permutations
of the particle coordinates. The Hamiltonian $\tilde H_N$ is a
symmetric, densely defined operator and by Lemma~\ref{lm:a1}, it
is bounded from below. The quadratic form \(q(\ph,\psi) =
\sprod{\ph}{\tilde H_N\psi}\) with domain $D=\D_N$ is therefore
bounded below and closable and hence the theory of the previous
section applies, once we have verified assumptions (i), (ii), and
(iii). The unique self-adjoint operator $H_N$ associated with the
closure of the quadratic form $q$ is the Friedrichs' extension of
$\tilde H_N$. The thresholds $\Sigma_R$ and $\Sigma$ associated
with the form $q$ are now given by
\begin{eqnarray*}
\Sigma_R(H_N) &=& \inf_{\ph\in \D_{N,R},\ \|\ph\|=1}\sprod{\ph}{\tilde H_N\ph}\\
\Sigma(H_N) &=& \lim_{R\to\infty} \Sigma_R(H_N)
\end{eqnarray*}
where \(\D_{N,R} := \{\ph\in \D_N: \ph(X)=0\ \text{if}\ |X|<
R\}\). The following theorem is a corollary of
Theorem~\ref{thm:main1}.

\begin{theorem}[Exponential decay in QED]\label{thm:main3}
  Assume Hypothesis (H1) is satisfied and let $H_N$ be the
  Friedrichs' extension of the symmetric operator \(\tilde
  H_N:\D_N\subset \H_N\to\H_N\) given by Eq.~\eqref{eq:ham}. If
  $\lambda$ and $\beta$ are real numbers with \(\lambda
  +\beta^2<\Sigma(H_N)\), then
  \[\Big\|e^{\beta|X|}E_{\lambda}(H_N)\Big\| <\infty.\]
\end{theorem}

\begin{proof}
It suffices to verify the assumptions (i), (ii) and (iii) in the
previous section. Suppose \(f\in C^{\infty}(\R^{3N})\) with
\(f,\nabla f\in L^{\infty}(\R^{3N})\), and \(f(X)=f(|X|)\). Then
\(f\D_N\subset \D_N\) is obvious from the definition of $\D_N$.
Property (ii) follows from
\begin{eqnarray*}
f(p_i+\sqrt{\alpha}A(x_i))^2 f & \leq &
2\|f\|_{\infty}^2(p_i+\sqrt{\alpha}A(x_i))^2 +
2\|\nabla_{x_i} f\|_{\infty}^2\\
f H_f f &\leq & \|f\|_{\infty}^2 H_f\\
fVf & \leq & \|f\|_{\infty}^2 V_{+},
\end{eqnarray*}
from Lemma~\ref{lm:a2} and Lemma~\ref{lm:a1}. The proof of (iii)
is a straightforward computation using that \(f^2 \tilde H_N +\tilde
H_N f^2 -2f \tilde H_N f = [[\tilde H_N,f],f]\).
\end{proof}

\emph{Remark.} The above theorem and its proof can easily be
generalized to systems of $N$ particles from $n\leq N$ species,
with different masses $m_i$, charges, and spins.
Theorem~\ref{thm:main3} then equally holds with the new norm
\(|X|=\big(\sum_{i=1}^N 2m_i x_i^2\big)^{1/2}\) in the factor
$e^{\beta|X|}$.

Our next goal is to establish a relation between $\Sigma(H_N)$
and spectral data of cluster Hamiltonians. To this end, we impose
the following additional assumption on $V$:
\begin{equation*}
  (H2) \left\{\begin{array}{l}
  V(X) = \sum_{i=1}^{N}v(x_i) + \sum_{i<j}w(x_i-x_j)\quad
  \text{where}\ v,w\in L^2_{\rm loc}(\R^3)\\
  \text{and}\ \lim_{|x|\to\infty}v(x)=0,\
  \lim_{|x|\to\infty}w(x)=0.
  \end{array}\right.
\end{equation*}

If the external potential $v$ is associated with a particle
sitting at the origin $x=0$, then these assumptions can be
understood as saying that the interaction between spatially
separated clusters of particles drops off to zero as the
inter-cluster distance increases to infinity. The limitation to
two-body forces in (H2) is not necessary.

\begin{theorem}[Equivalence of ionization thresholds]\label{thm:main2}
Assume (H1) and (H2), and let
\(E_{N-N'}=\inf\sigma(H_{N-N'})\), and
\(E_N^0=\inf\sigma(H_N^0)\) where all external potentials are
dropped in $H_N^0$. Then
\[\Sigma(H_N) = \min_{N'\geq 1}\{E_{N-N'} + E_{N'}^0\}.\]
\end{theorem}

The proof requires, in particular, localizing the field  energy in
neighborhoods of the electrons. In order to control the
localization errors which thereby arise we need an infrared cutoff
in the interaction. That is, we first prove the above theorem in
the case where all interactions of electrons with photons of
energy less than an arbitrary small, but positive constant $\mu$
have been dropped from $H_N$. The theorem then follows in the
limit $\mu\to 0$.


\section{IR-Cutoff Hamiltonians}\label{sec:proofs}

In this section we prove Theorem~\ref{thm:main2} by first
establishing an analogous results for Hamiltonians with an
infrared cutoff $\mu$ in the interaction. Theorem~\ref{thm:main2}
then follows in the limit $\mu\to 0$.

The infrared cutoff Hamiltonians $H_{N,\mu}$, $\mu>0$, are defined
in the same way as $H_N$ with the only difference that the vector
potential $A(x)$ and the magnetic field $B(x)$ in $H_{N}$ are
replaced by
\begin{equation*}
  A_{\mu}(x) =  \sum_{\lambda=1,2}\int_{\mu\leq |k|\leq
  \Lambda} \frac{1}{\sqrt{|k|}}\eps_{\lambda}(k) \Big[ e^{ik\cdot
  x} a_{\lambda}(k) +e^{-ik\cdot x} a_{\lambda}^{*}(k) \Big]d^3k
\end{equation*}
and $B_{\mu}(x)=\curl A_{\mu}(x)$. To separate the soft,
non-interacting photons from the interacting ones we use that
$\F_f$ is isomorphic to \(\F_i\otimes\F_s\) where $\F_i$ and
$\F_s$ denote the bosonic Fock spaces over $L^2(|k|\geq \mu)$ and
$L^2(|k|< \mu)$ respectively. Let $\H_i=\wedge^N
L^2(\R^3;\C^2) \otimes\F_i$. Then the Hamilton operator can be
written as
\begin{equation}\label{eq:decomp}
  H_{N,\mu} = H_{\mu}^i\otimes 1 + 1\otimes H_f^s\qquad
  \text{on}\quad \H=\H_i\otimes \F_s
\end{equation}
if we identify $\F$ with \(\F_i\otimes\F_s\). Let $\F_{s,n}$
denote the $n$-boson subspace of $\F_s$ and let $\Omega_s$ be the
vacuum of $\F_s$. Then \eqref{eq:decomp} and the positivity of
$H_f^s=H_f\restricted \F_s$ imply that
\begin{equation}\label{eq:nosoft}
\begin{aligned}
\inf\sigma(H_{N,\mu}) &= \inf_{n\geq 0
}\big(\inf\sigma(H_{N,\mu}\restricted
\H_i\otimes \F_{s,n})\big)\\
 &= \inf\sigma(H_{N,\mu}\restricted \H_i\otimes [\Omega_s]),
\end{aligned}
\end{equation}
where $[\Omega_s]$ is the space spanned by $\Omega_s$.
This will allow us to drop the soft bosons in all approximate
energy minimizers.

\begin{lemma}\label{lm:H-Hir}
There exists a constant $C_{N,\Lambda}$, depending on \(\Lambda,
N, g\) and $\alpha$, such that
\begin{equation*}
\pm (H_N-H_{N,\mu}) \leq
\mu^{1/2}C_{N,\Lambda}\left\{\sum_{i=1}^{N}p_i^2+H_f+1\right\}\qquad\text{for}\quad
0\leq \mu\leq 1.
\end{equation*}
\end{lemma}

\begin{proof}
By definition of $H_N$ and $H_{N,\mu}$,
\begin{eqnarray*}
H_N-H_{N,\mu} &=& \sum_{i=1}^N 2\sqrt{\alpha} p_i\cdot
\big(A(x_i)-A_{\mu}(x_i)\big) +
\alpha\big(A(x_i)-A_{\mu}(x_i)\big)\cdot\big(A(x_i)+A_{\mu}(x_i)\big)\\
& & + \frac{g}{2}\sqrt{\alpha} \sigma\cdot
\big(B(x_i)-B_{\mu}(x_i)\big)
\end{eqnarray*}
where we used that $A(x)$ and $A_{\mu}(x)$ commute. The
differences \(A(x)-A_{\mu}(x)\) and \(B(x)-B_{\mu}(x)\) can be
seen as a vector potential and a magnetic field with an
\emph{ultraviolett} cutoff $\mu$. Hence the lemma follows from
Lemma~\ref{lm:a2} with $\Lambda=\mu$ and \(\eps=\mu^{1/2}\).
\end{proof}

\begin{lemma}\label{lm:T-Tir}
\begin{itemize}
\item[(i)] $\Sigma(H)<\infty$ if and only if
  $\Sigma(H_{\mu})<\infty$ and in this case there exists a constant
  $C_{\Lambda}$ depending on the parameters $\Lambda, N, \alpha$, and
  $g$, such that
  \[|\Sigma(H_{N,\mu})-\Sigma(H_N)| \leq C_{\Lambda} \mu^{1/2},
  \qquad\text{if}\ \mu\leq 1.\]
\item[(ii)] There exists a constant $C_{\Lambda}$ depending on the
  parameters $\Lambda, N, \alpha$, and
$g$, such that
  \[|\tau(H_{N,\mu}) - \tau(H_N)| \leq C_{\Lambda} \mu^{1/2},\qquad\text{if}\ \mu\leq1.\]
\end{itemize}
\end{lemma}

\begin{proof}
By Lemma~\ref{lm:H-Hir} and Lemma~\ref{lm:a1}, there exist
constants $C$ and $D$, independent of $\mu$, such that
\begin{equation*}
  H_{N,\mu} \leq H_N + \mu^{1/2}(C H_N+D)\qquad \text{for}\quad \mu\leq 1.
\end{equation*}
It follows that
\begin{equation*}
  \Sigma(H_{N,\mu}) \leq \Sigma(H_N) + \mu^{1/2}(C \Sigma(H_N) +D)\qquad \text{for}\quad \mu\leq 1
\end{equation*}
and, in particular, that $\Sigma(H_{N,\mu})<\infty$ if $\Sigma(H_{N})<\infty$.
Since the roles of $H_{N,\mu}$ and $H_{N}$ are interchangeable, (i) follows.
The
proof of (ii) is similar.
\end{proof}

\begin{theorem}\label{thm:with-ir}
Suppose assumptions (H1) and (H2) on $V$ are satisfied. Then
\[  \Sigma(H_{N,\mu}) = \tau(H_{N,\mu})\qquad \text{for all}\ \mu>0.\]
\end{theorem}
In conjunction with Lemma~\ref{lm:T-Tir}, this theorem proves
Theorem~\ref{thm:main2}.

\begin{proof}[Proof of \(\Sigma(H_{N,\mu}) \geq \tau(H_{N,\mu})\).]
The key element for this proof is Theorem~\ref{thm:GLL}, whose
long proof is given in \cite{GLL}. Here we merely show how
\(\Sigma(H_{N,\mu}) \geq \tau(H_{N,\mu})\) follows from
Theorem~\ref{thm:GLL}. We may certainly assume that
\(\Sigma(H_{N,\mu})<\infty\). By the argument \eqref{eq:nosoft} we
may restrict $H_{N,\mu}$ to \(\H_i\otimes [\Omega_s]\) for the
computation of $\Sigma_R(H_{N,\mu})$. By Lemma~\ref{lm:a1}
\begin{equation*}
  N_f \leq   \frac{1}{\mu} H_f \leq
  \frac{1}{\mu} (2 H_{N,\mu}+D)
\qquad{on}\quad \H_i\otimes [\Omega_s]
\end{equation*}
and hence by Theorem~\ref{thm:GLL},
\begin{equation*}
  H_{N,\mu} \geq \tau(H_{N,\mu}) - o(R^{0}) (H_{N,\mu} + C)
  \qquad{on}\quad \D_{N,R}\cap (\H_i\otimes [\Omega_s]).
\end{equation*}
It follows that
\begin{equation*}
  \Sigma_{R}(H_{N,\mu}) \geq \tau(H_{N,\mu}) - o(R^{0}) (\Sigma_R(H_{N,\mu}) + C)
\end{equation*}
and the desired result is obtained in the limit $R\to \infty$.
\end{proof}


An important role in the following proof is played by the
identification operator \(I:\F\otimes\F\to\F\) which collects all
photons in the first and second factor of $\F\otimes\F$, and
gathers them in a single Fock space. For the precise definition of
$I$, and for notations in the following proof that have not yet
been introduced, see Appendix~\ref{sec:Fock}.

\begin{proof}[Proof of \(\Sigma(H_{N,\mu}) \leq \tau(H_{N,\mu})\).]
In the following the subindex $\mu$ is dropped. We need to show that
\[ \lim_{R\to\infty} \Sigma_R(H_N) \leq E_{N-N'} + E_{N'}^0\]
for all $N'\geq 1$. The strategy is as follows. First we construct
approximate minimizers $\ph_0$ and $\ph_{\infty}$ of $H_{N-N'}$
and $H_{N'}^0$ respectively, with the property that the electrons
and the photons described by $\ph_0$ and $\ph_{\infty}$ are
compactly supported. Then, by a translation \(\ph_{\infty}\to
T_R\ph_{\infty}\) of both the electrons and the photons in
$\ph_{\infty}$ we may achieve (ignoring the Pauli principle) that
\[ \psi_R = I(\ph_0\otimes T_R\ph_{\infty}) \in \D_{N,R},\qquad\text{and}\ \|\psi_R\|=1,  \]
where $T_R\ph_{\infty}$ is still an approximate minimizer of
$H_{N'}^0$ by the translation invariance of this Hamiltonian.

Second we show that
\[  \sprod{\psi_R}{H_N\psi_R} \leq \sprod{\ph_0}{H_{N-N'}\ph_0} +
\sprod{\ph_{\infty}}{H_{N'}^0\ph_{\infty}} + o(R^0)\qquad R\to
\infty\] which concludes the proof. To incorporate the Pauli
principle one needs to anti-symmetrize \(I(\ph_0\otimes
T_R\ph_{\infty})\) with respect to the $N$ electron variables
\((x_i,s_i)\in\R^3\times\{1,2\}\), $i=1,\ldots,N$. After
normalization, this will lead to the same value for the energy
\(\sprod{\psi_R}{H_N\psi_R}\) as without anti-symmetrization,
because the electrons in $\ph_0$ and $T_R\ph_{\infty}$ are
disjointly supported and the Hamiltonian is local. Therefore we
don't need to anti-symmetrize.

Let $\eps>0$ be given and fixed in the following three steps, and
let $y$ denote the position operator $y=i\nabla_k$ in the
one-photon Hilbert space. For simplicity the irrelevant
parameters $\alpha$ and $g$ are dropped henceforth.

{\bf Step 1.} Given $\eps>0$ there are normalized states
$\ph_0\in \D_{N-N'}$ and $\ph_{\infty}\in \D_{N'}$ such that
\begin{itemize}
\item[(i)] \(
\sprod{\ph_0}{H_{N-N'}\ph_0} < E_{N-N'} +
  \eps/2\quad\text{and}\quad
  \sprod{\ph_{\infty}}{H_{N'}^0\ph_{\infty}} < E_{N'}^0 + \eps/2 \).
\item[(ii)] Both \(\sprod{\ph_0}{N_f\ph_0}\) and
\(\sprod{\ph_{\infty}}{N_f\ph_{\infty}}\) are finite and bounded
by a constant that is independent of $\eps>0$.
\item[(iii)] $\ph_0$ and $\ph_{\infty}$ have compact support as
functions of the electronic configurations $X_{N-N'}\in
\R^{3(N-N')}$ and $X_{N'}\in \R^{3N'}$ respectively.
\item[(iv)] There exists an $R_0$ such that
\[\ph_0=\Gamma(\chi_{R_0})\ph_0,\quad \ph_{\infty}=\Gamma(\chi_{R_0})\ph_{\infty}\]
where \(\chi_{R_0}\) is the characteristic function of the ball
\(\{y\in \R^3:\, |y|< R_0\}.\)
\end{itemize}

\emph{Proof of Step 1.} The properties of the Hamiltonians that
are relevant, are shared by $H_{N-N'}$ and $H_{N'}^0$. So it
suffices to prove existence of $\ph_0$. Let \(H_0:= H_{N-N'}\)
and \(E_0:= E_{N-N'}\) for short. Let $\chi_P$ be the operator
of multiplication with $\chi(|X|/P)$ on $\H_{N-N'}$ where
\(\chi\in C^{\infty}(\R_{+})\), $\chi(t)=1$ for $t\leq 1$,
$\chi(t)=0$ for $t\geq 2$ and $0\leq \chi\leq 1$. Let $j_R$ be
the operator of multiplication with $\chi(|y|/R)$ on
$L^2(\R^3,dk)$. Existence of $\ph_0$ with property (i) and (ii)
follows from the fact that $\D_{N-N'}$ is a form core of $H_0$, the
argument \eqref{eq:nosoft}, and Lemma~\ref{lm:a1}. If we now show
that
\begin{eqnarray}
\sprod{\chi_P\ph_0}{(H_0-E_0)\chi_P\ph_0}
&\stackrel{P\to\infty}{\longrightarrow}&
\sprod{\ph_0}{(H_0-E_0)\ph_0}\label{eq:step1a}\\
\sprod{\Gamma(j_R)\chi_P\ph_0}{(H_0-E_0)\Gamma(j_R)\chi_P\ph_0}
&\stackrel{R\to\infty}{\longrightarrow}&
\sprod{\chi_P\ph_0}{(H_0-E_0)\chi_P\ph_0}\label{eq:step1b}
\end{eqnarray}
then (iii), and (iv) will follow, because, by the strong
convergence $\chi_P\to 1$ and $\Gamma(j_R)\to 1$ the norm
$\|\Gamma(j_R)\chi_P\ph_0\|$ is close to $1$ for large $P$ and
large $R$.

Properties \eqref{eq:step1a} and \eqref{eq:step1b} follow from
\begin{eqnarray}
\lim_{P\to\infty}[H_0,\chi_P]\ph_0 &=& 0\label{eq:step1d}\\
\lim_{R\to\infty}(N_f+1)^{-1/2}[H_0,\Gamma(j_R)]\chi_P\ph_0 &=&
0\label{eq:step1e}
\end{eqnarray}
(to be proven shortly) by commuting the operators $\chi_P$ and
$\Gamma(j_R)$ through $H_0-E_0$ and using (ii) and that
\(s-\lim_{P\to\infty}\chi_P^2=1\) and
\(s-\lim_{R\to\infty}\Gamma(j_R)^2=1\). Note that
\(\Gamma(j_R)\chi_P \D_{N-N'}\subset \D_{N-N'}\).

Equation~\eqref{eq:step1d} follows from
\[ [H_0,\chi_P] = \sum_{j=1}^{N-N'}(-2i)\nabla_{x_j}\chi_{P}\cdot(p_j+A(x_j)) -
\Delta_{x_j}\chi_{P}\] using $\nabla_{x_j}\chi_P= O(P^{-1})$,
$\Delta_{x_j}\chi_{P}= O(P^{-2})$ and Lemma~\ref{lm:a1}.

To prove \eqref{eq:step1e} we write the commutator as
\begin{equation}\label{eq:step1f}
\begin{split}
[ H_0,\Gamma(j_R) ] =& \sum_{i=1}^{N-N'}
\Big\{(p_i+A(x_i))[A(x_i),\Gamma(j_R)] +
[A(x_i),\Gamma(j_R)](p_i+A(x_i))\\
 & \qquad+ [\sigma_i\cdot B(x_i),\Gamma(j_R)] +
[H_f,\Gamma(j_R)]\Big\}
\end{split}
\end{equation}
Using that \(H_f=\dGamma(|k|)\), the last term in \eqref{eq:step1f}
restricted to \(\otimes_s^n L^2(\R^3)\) is given by \([H_f,\Gamma(j_R)]
= \sum_{l=1}^n j_R\otimes\ldots\otimes[|k|,j_R]\ldots\otimes j_R\),
the commutator being the $l$th factor.
Since \(\|[|k|,j_R]\|=O(R^{-1})\) it follows that
\(\|(N+1)^{-1/2}[H_f,\Gamma(j_R)](N+1)^{-1/2}\|=O(R^{-1})\), and
hence, by (ii), that the contribution due to $H_f$ is of order
$R^{-1}$. To deal with the first two terms in \eqref{eq:step1f}
note that, by \eqref{geq1} and \eqref{geq2},
\begin{equation*}
  [A(x_i),\Gamma(j_R)] = a^*((1-j_R)G_{x_i})\Gamma(j_R) -
  \Gamma(j_R)a((1-j_R)G_{x_i})
\end{equation*}
where
\[ \|a^{\sharp}((j_R-1)G_{x_i})\chi_P(N+1)^{-1/2}\|\leq
\sup_{|x_i|\leq 2P}\|(j_R-1)G_{x_i}\|\to 0, \qquad \text{as}\quad
R\to \infty.\] It follows that the terms in \eqref{eq:step1f}
which are quadratic in $A(x_i)$ give vanishing contributions, as
the factors $A(x_i)$ outside the commutators can be controlled by
$(N+1)^{-1/2}$. To show that the terms in \eqref{eq:step1f} with
an operator $p_i$ vanish in the limit $R\to\infty$ it suffices to
add to the above arguments that
\(p_i[A(x_i),\Gamma(j_R)]=[A(x_i),\Gamma(j_R)]p_i\) because $p_i$
commutes with $A(x_i)$ and $\Gamma(j_R)$, that \(p_i\chi_P=\chi_P
p_i -i \nabla_i \chi_P\) and that \(\|p_i\ph_0\|<\infty\) by
Lemma~\ref{lm:a1}. The term involving $B(x_i)$ is dealt with
similarly.


{\bf Step 2.} Let $\eps,\ \ph_0$, and $\ph_{\infty}$ be as in Step
1. Pick $R_0$ so large that, with $\chi_{R_0}$ is as in Step 1 (iv), \(\ph_0=\Gamma(\chi_{R_0})\ph_0\),
\(\ph_{\infty}= \Gamma(\chi_{R_0})\ph_{\infty}\),
\(\ph_{0}(X_{N-N'})=0\) if
$|X_{N-N'}|>R_0$ and \(\ph_{\infty}(X_{N'})=0\) if
$|X_{N'}|>R_0$. Let $R\geq R_0$ and pick a vector $d\in \R^3$
with $|d|=3$. Let $T_R:\H_{N-N'}\to \H_{N-N'}$ be the translation
\begin{equation*}
  T_R = \exp\left(-i Rd\cdot \Big\{\sum_{i=1}^{N'}p_i+P_f\Big\}\right)
\end{equation*}
where $P_f=\dGamma(k)$ is the total momentum operator of the
photons. Then
\begin{eqnarray*}
\sprod{T_R\ph_{\infty}}{H_{N'}^0 T_R\ph_{\infty}} &=&
\sprod{\ph_{\infty}}{H_{N'}^0 \ph_{\infty}},\\
\psi_R := I(\ph_0\otimes T_R \ph_{\infty}) &\in & \D_{N,2R}.
\end{eqnarray*}


{\bf Step 3.} If
$R\geq R_0$ then $\|\psi_R\|=1$ and
\begin{equation*}
  \sprod{\psi_R}{H_N\psi_R} = \sprod{\ph_0}{H_{N-N'}\ph_0} +
\sprod{\ph_{\infty}}{H_{N'}^0\ph_{\infty}} + o(R^0),\quad R\to
\infty.
\end{equation*}
In particular \(\Sigma_R(H_N)\leq E_{N-N'} + E_{N'}^0+2\eps\) for
all $R$, which proves the theorem.

\emph{Proof of Step 3.} By construction of $\ph_0$ and
$T_R\ph_{\infty}$ the photons in these states have disjoint
support if $R\geq R_0$. Therefore
\begin{eqnarray*}
\sprod{\psi_R}{\psi_R} &=& \sprod{I(\ph_0\otimes
T_R\ph_{\infty})}{I(\ph_0\otimes T_R\ph_{\infty})}\\
&=& \sprod{\ph_0\otimes
T_R\ph_{\infty}}{\ph_0\otimes T_R\ph_{\infty}}\\
&=& \sprod{\ph_0}{\ph_0}\sprod{\ph_{\infty}}{\ph_{\infty}} = 1.
\end{eqnarray*}
In the following this property of $I$, that it acts like an
isometry on product states with photons supported in \(\{|y|\leq
R_0\}\) and \(\{|y-Rd|\leq R_0\}\) respectively, will be used
repeatedly and tacitly.

Writing \(H_f=\sum_{\lambda=1,2}\int |k|
a^{*}_{\lambda}(k)a_{\lambda}(k)d^3k\) and using \eqref{eq:aI}
one gets
\begin{eqnarray*}
\sprod{\psi_R}{H_f\psi_R} &=&
\sprod{\ph_0}{H_f\ph_0}+\sprod{T_R\ph_{\infty}}{H_{f}T_R\ph_{\infty}}\\
&& + 2\Rea \sum_{\lambda=1,2}\int |k|
\sprod{a_{\lambda}(k)\ph_0}{\ph_0}
\sprod{\ph_{\infty}}{a_{\lambda}(k)\ph_{\infty}}e^{iRd\cdot k}d^3k
\end{eqnarray*}
where \(T_R^{*}a(k) T_R = e^{iRd\cdot k}a(k)\) was also used. The
third term converges to zero as $R\to\infty$ by the
Riemann-Lebesgue lemma, because the integrand is in
\(L^1(\R^3,\C^2)\).

Since the distance of the electrons described by $T_R\ph_{\infty}$
to the origin and to the electrons in $\ph_{0}$ is bounded below
by \(3R-3R_0\), we have, by assumption (H2), that
\begin{equation*}
  \sprod{\psi_R}{V_N\psi_R} = \sprod{\ph_0}{V_{N-N'}\ph_0} +
  \sum_{i<j}\sprod{T_R\ph_{\infty}}{w(x_i-x_j) T_R\ph_{\infty}} + o(R^0),\qquad
  (R\to\infty),
\end{equation*}
as desired. Next we compare
\[ \sum_{j=1}^N\sprod{\psi_R}{(p_j+A(x_j))^2\psi_R}\]
with
  \[ \sum_{j\leq N-N'}\sprod{\ph_0}{(p_j+A(x_j))^2\ph_0} +
  \sum_{j> N-N'}\sprod{T_R\ph_{\infty}}{(p_j+A(x_j))^2 T_R
  \ph_{\infty}}.\]
To this end we write \(A(x_j)=
a(G_{x_j})+a^{*}(G_{x_j})\) and use that
\begin{eqnarray*}
(p_j+A(x_j))^2 &=& p_j^2 + 2p_j\cdot a(G_{x_j}) +
2a^{*}(G_{x_j})\cdot p_j \\
&& + a(G_{x_j})^2 + a^{*}(G_{x_j})^2 +
2a^{*}(G_{x_j})a(G_{x_j})+\|G_{x_j}\|^2.
\end{eqnarray*}
Let $j\leq N-N'$, then using \eqref{eq:aI} and again disjointness
of the supports of the photons in $\ph_0$ and $T_R\ph_{\infty}$, one finds that
\begin{eqnarray*}
\sprod{\psi_R}{(p_j+A(x_j))^2\psi_R} &=&
\sprod{\ph_0}{(p_j+A(x_j))^2\ph_0}\\
&& + 2\sprod{\ph_0\otimes T_R\ph_{\infty}}{p_j\ph_0\otimes
a(G_{x_j})T_R\ph_{\infty}}\ +\ \text{h.c.}\\
&& + 2\sprod{\ph_0\otimes T_R\ph_{\infty}}{a(G_{x_j})\ph_0\otimes
a(G_{x_j})T_R\ph_{\infty}}\ +\ \text{h.c.}\\
&& + \sprod{\ph_0\otimes T_R\ph_{\infty}}{\ph_0\otimes
a(G_{x_j})^2T_R\ph_{\infty}}\ +\ \text{h.c.}\\
&& + \sprod{a(G_{x_j})\ph_0\otimes T_R\ph_{\infty}}{\ph_0\otimes
a(G_{x_j})T_R\ph_{\infty}}\ +\ \text{h.c.}\\
&& + \sprod{\ph_0\otimes a(G_{x_j})T_R\ph_{\infty}}{\ph_0\otimes
a(G_{x_j})T_R\ph_{\infty}}.
\end{eqnarray*}
All terms except the
first one vanish in the limit $R\to\infty$. In fact,
\begin{eqnarray*}
a(G_{x_j})T_R\ph_{\infty} &=& T_R
a(G_{x_j-Rd})\Gamma(\chi_{R_0})\ph_{\infty}\\
 &=& T_R \Gamma(\chi_{R_0})a(\chi_{R_0}G_{x_j-Rd})\ph_{\infty},
\end{eqnarray*}
and since \(|x_j|\leq R_0\) if \(\ph_0(x_1,\ldots,x_{N-N'})\neq
0\), we can multiply this in all the above terms with
\(\chi_{R_0}(x_j)\). But then, by \eqref{eq:aN_bound} and using
the notation
\(G_{\lambda}(k)=|k|^{-1/2}\eps_{\lambda}(k)\chi_{\Lambda}(k)\)
\begin{multline}\label{eq:step3a}
  \|\chi_{R_0}(x_j)a(\chi_{R_0}G_{x_j-Rd})(N_f+1)^{-1/2}\|^2\\ \leq
  \sup_{|x_j|\leq
R_0} \sum_{\lambda=1,2}\int_{|y|\leq R_0} |\hat{G}_{\lambda}(x_j-Rd-y)|^2 dy \to 0\qquad (R\to \infty).
\end{multline}

The case where $j>N-N'$ is dealt with similarly. The only
difference there is that \(|x_j-Rd|\leq R_0\) in the support of
$T_R\ph_{\infty}$ and the photons in $\ph_0$ have support in
$|y|\leq R_0$. Hence \eqref{eq:step3a} will be replaced by
\begin{multline*}
  \|\chi_{R_0}(x_j-Rd)a(\chi_{R_0}G_{x_j})(N_f+1)^{-1/2}\|^2\\ \leq \sup_{|x_j-Rd|\leq
  R_0} \sum_{\lambda=1,2}\int_{|y|\leq R_0} |\hat{G}_{\lambda}(x_j-y)|^2 dy \to 0\qquad (R\to \infty).
\end{multline*}
The terms involving $B(x_i)$ are dealt with similarly.
\end{proof}


\appendix

\section{Important Estimates}

\begin{lemma}\label{lm:a2}
For all $\Lambda\geq 0$, $\eps>0$ and all $x\in\R^3$,
\begin{eqnarray*}
A(x)^2 &\leq & 32\pi \Lambda (H_f + \Lambda/4),\\
\pm \sigma\cdot B(x) & \leq & \eps H_f
+\frac{8\pi}{\eps}\Lambda^3.
\end{eqnarray*}
\end{lemma}

For the proof see \cite{GLL}. This lemma holds equally for
$A_{\mu}(x)$ and $B_{\mu}(x)$ with $\mu>0$.

\begin{lemma}\label{lm:a1}
Let $C=1+32 \pi \alpha N\Lambda$ and $D=8\pi\alpha N\Lambda$.
Then, for all $\mu\geq 0$,
\begin{equation*}
\sum_{i=1}^N p_i^2 \leq  C \left\{\sum_{i=1}^N
(p_i+\sqrt{\alpha}A_{\mu}(x_i))^2 + H_f \right\} + D.
\end{equation*}
Furthermore, if \(V_{-}\leq \eps p^2+C_{\eps}\) for all $\eps>0$,
then there exist constants $D(\eps)$, depending on
\(\alpha,g,N,\Lambda\) and $\eps$, but not on $\mu$, such that
\begin{equation*}
  \left\{\sum_{i=1}^N (p_i+\sqrt{\alpha}A_{\mu}(x_i))^2 + V_{+} + H_f\right\}
  \leq (1+\eps) H_{N,\mu}+D(\eps).
\end{equation*}
\end{lemma}

\begin{proof}
The first part follows from \(p_i^2\leq
2(p_i+\sqrt{\alpha}A_{\mu}(x_i))^2+2\alpha A_{\mu}(x_i)^2\) and
Lemma~\ref{lm:a2}. The second bound follows from the first and
Lemma~\ref{lm:a2}.
\end{proof}

\begin{theorem}\label{thm:GLL}
Suppose the negative parts $v_{-}$ and $w_{-}$ of the external
potential $v$ and the two-particle interaction $w$ as functions
in $\R^3$ drop off to zero as \(|x|\to\infty\). Then for all
values of the parameters $N, \Lambda, \alpha, g$ and $\mu\geq 0$,
there exists a functions $f(R)$  and a constant $C$, depending on
these parameters, such that
\begin{equation*}
  H_{N,\mu} \geq \tau(H_{N,\mu}) - f(R)(H_{N,\mu}+N_f+C)
  \qquad\text{on}\ \D_{N,R}
\end{equation*}
where \(\lim_{R\to\infty}f(R)=0\). Here \(\tau(H_{N,\mu}) =
\inf_{N'\geq 1}[\inf\sigma(H_{N-N',\mu}) +
\inf\sigma(H_{N,\mu}^0)]\).
\end{theorem}

This theorem is a variant of Corollary A.2 in \cite{GLL}, where we
used the positivity of the photon mass to estimate $N_{f}$ in
terms of $H_{N}$. Thus the error term in Corollary A.2 of
\cite{GLL} depends on the photon mass. This was overlooked in
\cite{GLL} leaving a gap in the proof. Theorem~\ref{thm:main2}
combined with Theorem~5.1 in \cite{GLL} closes the gap.

\section{Fock Space and Second Quantization}
\label{sec:Fock}

Let $\h$ be a complex Hilbert space, and let $\otimes_s^n\h$
denote the symmetric tensor product of $n$ copies of $\h$. Then
the bosonic Fock space over $\h$:
\[  \F=\F(\h)=\oplus_{n\geq0}\otimes_s^n\h \]
is the space of sequences \(\ph=(\ph_n)_{n\geq 0}\), with
$\ph_0\in \C$, \(\ph_n\in \otimes_s^n\h\), and with an inner
product defined by
\[ \sprod{\ph}{\psi} := \sum_{n\geq 0} ( \ph_n  , \psi_n ), \]
where \(( \ph_n,\psi_n)\) denotes the inner product of
\(\otimes^n_s \h\). The vector \(\Omega=(1,0,\ldots)\in\F\) is
called the vacuum. By $\Fin\subset\F$ we denote the dense
subspace of vectors $\ph$ for which $\ph_n=0$, for all but
finitely many $n$. The number operator $N_f$ in $\F$ is defined by
\((N_f\ph)_n=n\ph_n\).

\subsection{Creation- and Annihilation Operators}

The creation operator $a^*(h)$, $h\in\h$, on $\Fin\subset \F$ is
defined by $(a^*(h)\ph)_0 =0$ and
\[ (a^*(h)\ph)_n = \sqrt{n}\,S_n(h\otimes \ph_{n-1})\]
where \(S_n\in \L(\otimes^n\h)\) denotes the orthogonal projection
onto the symmetric subspace \(\otimes_s^n\h\subset \otimes^n\h\).
The annihilation operator $a(h)$ is the adjoint of $a^*(h)$
restricted to $\Fin$. Creation- and annihilation operators satisfy
the canonical commutation relations (CCR)
\begin{equation*}
[a(g),a^{*}(h)] = (g,h),\hspace{3em} [a^{\#}(g),a^{\#}(h)] =0.
\end{equation*}
In particular \([a(h),a^{*}(h)] = \|h\|^2\). From the definition
of $a^*(h)$ it is easy to see that
\begin{equation}\label{eq:aN_bound}
\|a^{\#}(h)(N+1)^{-1/2}\| \leq \|h\|.
\end{equation}

In the case where $\h$ is the one-photon Hilbert space,
\(L^2(\R^3;\C^2)\), the annihilation and creation operators can be
expressed in terms of the operator-valued distributions
$a_{\lambda}(k)$ and $a_{\lambda}^{*}(k)$ by
\begin{eqnarray*}
a(h) &=& \sum_{\lambda=1,2} \int
\overline{h_{\lambda}(k)}a_{\lambda}(k)\, d^3k\\
a^{*}(h) &=& \sum_{\lambda=1,2} \int
h_{\lambda}(k)a_{\lambda}^{*}(k)\, d^3k.
\end{eqnarray*}
Setting
\(G_{x,\lambda}(k)=|k|^{-1/2}\eps_{\lambda}(k)\chi_{\{|k|\leq
\Lambda\}}e^{-ik\cdot x}\), the quantized vector potential $A(x)$
can be defined as \(A(x)=a(G_x)+ a^{*}(G_x)\).

\subsection{Second Quantization}

Suppose $b$ is a bounded operator on $\h$ and $\|b\|\leq 1$. The
operator $\Gamma(b)\ :\ \F(\h)\rightarrow\F(\h)$ is defined by
\begin{eqnarray*}
  \Gamma(b) \Omega &=& \Omega\\
  \Gamma(b)\restricted\otimes_s^n \h &=& b\otimes\ldots\otimes b.
\end{eqnarray*}
Clearly \(\|\Gamma(b)\|\leq 1\). From the definition of $a^{*}(h)$ it
easily follows that
\begin{eqnarray}
\Gamma(b)a^{*}(h) & =& a^{*}(bh)\Gamma(b) \label{geq1}\\
\Gamma(b)a(b^*h) &=& a(h)\Gamma(b),\label{geq2}
\end{eqnarray}
and hence that \(\Gamma(b)a(h) = a(bh)\Gamma(b)\) if \(b^{*}b=1\).

If \(b: D(b)\subset\H\to \H\) is self-adjoint, then $\dGamma(b)$ in $\F(\h)$ is defined by
\begin{eqnarray*}
\dGamma(b)\Omega &=& 0\\
\dGamma(b)\restricted\otimes_s^n D(b) &=& \sum_{j=1}^n
(\underbrace{1\otimes\ldots 1}_{j-1}\otimes b\otimes
\underbrace{1\otimes\ldots 1}_{n-j})
\end{eqnarray*}
and by linear extension. $\dGamma(b)$ is essentially self-adjoint
and, denoting the closure by $\dGamma(b)$ as well,
\(\Gamma(e^{ib})=e^{i\dGamma(b)}\). One example is the number
operator $N_f=\dGamma(1)$, another one, for \(\h=L^2(\R^3;\C^2)\),
is the field energy
\begin{equation*}
  H_f = \dGamma(|k|) = \sum_{\lambda=1,2}\int
  |k|a^{*}_{\lambda}(k)a_{\lambda}(k)\, d^3k.
\end{equation*}

\subsection{The Identification Operator $I:\F\otimes\F\to\F$}

In the proof of Theorem~\ref{thm:with-ir} an important role is
played by the identification operator \( I:\F\otimes\F\to \F\)
defined by
\begin{align*}
I (\ph\otimes\Omega) &= \ph\\ I \ph\otimes a^*(h_1)\cdots
a^*(h_n)\Omega &= a^*(h_1)\cdots a^*(h_n)\ph,
\hspace{3em}\ph\in\Fin,
\end{align*}
and linear extension to $\Fin\otimes\Fin$. This operator is
unbounded. We often use the commutation relation
\begin{equation}\label{eq:aI}
  a(h) I = I(a(h)\otimes 1 + 1\otimes a(h)),
\end{equation}
which is in contrast to \(a^*(h)I=I(a^*(h)\otimes 1)
=I(1\otimes a^*(h))\).\\

\noindent {\bf Acknowledgements.} Part of this paper grew out of
the argument that closes a gap in \cite{GLL}. I am indebted to
Jean-Marie Barbaroux for pointing out the gap. This work was
completed when the author visited ETH Z\"urich and the University
of Mainz in the Summer of 2002. It is a pleasure to thank the
respective hosts, J\"urg Fr\"ohlich and Volker Bach, for their
hospitality and for useful discussions. I also thank Oliver Matte
for his careful proofreading.


\end{document}